# Graphene-based Distributed 3D Sensing Electrodes for Mapping Spatiotemporal Auricular Physiological Signals

Qingyun Huang[1,7], Cong Wu[1,7], Senlin Hou[1], Hui Sun[1], Kuanming Yao[2], Junhui Law[3], Mingxiao Yang[4], Vellaisamy A. L. Roy[5], Xinge Yu[2], Ho-yin Chan[1*], Lixing Lao[6], Yu Sun[3], and Wen Jung Li[1*]

[1] Department of Mechanical Engineering, City University of Hong Kong, Hong Kong, China
[2] Department of Biomedical Engineering, City University of Hong Kong, Hong Kong, China
[3] Department of Mechanical and Industrial Engineering, University of Toronto, Canada
[4] Bendheim Integrative Medicine Center, Memorial Sloan Kettering Cancer Center, USA
[5] James Watt School of Engineering, University of Glasgow, UK
[6] Virginia University of Integrative Medicine, USA
[7] These authors contributed equally: Qingyun Huang & Cong Wu
* Email: hoychan@cityu.edu.hk & wenjli@cityu.edu.hk

## ABSTRACT

Underneath the ear skin there are richly branching vascular and neural networks that ultimately connecting to our heart and brain. Hence, the three-dimensional (3D) mapping of auricular electrophysiological signals could provide a new perspective for biomedical studies such as diagnosis of cardiovascular diseases and neurological disorders. However, it is still extremely challenging for current sensing techniques to cover the entire ultra-curved auricle. Here, we report a graphene-based ear-conformable sensing device with embedded and distributed 3D electrodes which enable full-auricle physiological monitoring. The sensing device, which incorporates programable 3D electrode thread array and personalized auricular mold, has 3D-conformable sensing interfaces with curved auricular skin, and was developed using one-step multi-material 3D-printing process. As a proof-of-concept, spatiotemporal auricular electrical skin resistance (AESR) mapping was demonstrated. For the first time, 3D AESR contours were generated and human subject-specific AESR distributions among a population were observed. From the data of 17 volunteers, the auricular region-specific AESR changes after cycling exercise were observed in 98% of the tests and were validated via machine learning techniques. Correlations of AESR with heart rate and blood pressure were also studied using statistical analysis. This 3D electronic platform and AESR-based new biometrical findings show promising biomedical applications.

## INTRODUCTION

Intelligent wearable electronics have become widespread with the advancements in material science, fabrication technologies, and data science, for diverse healthcare applications involving physiological signal monitoring[1-10], including electrocardiography (ECG), electroencephalography (EEG), heart rate (HR), blood pressure (BP), body temperature, etc. These vital physiological signs carrying various biophysical or biochemical information from human bodies are critical for pursuing long-term and large-scale monitoring for clinical diagnosis of diverse diseases[11,12]. Many electronic devices worn on wrists, fingers, head, or legs recorded signals at single locations; however, signals measured at various spatial locations could provide additional dimension of crucial physiological information. For instance, full-scalp EEG monitoring with high-density electrode arrays enables recording of electrical activity in multiple positions across the brain with complementary information provided during functional magnetic resonance imaging[13-15], and spatiotemporal cardiac measurements of ECG, pH and temperature by integumentary membranes with conformable electrode arrays enables 3D mapping of epicardial signals[16].

Human ears also provide diverse physiological signals for health monitoring[17-24], such as oxygen saturation, pulse, EEG, body temperature and etc. Underneath the auricular skin, a complicated nerve network involving branches of the greater auricular nerve, lesser occipital nerve, auriculotemporal nerve, facial nerve, and vagus nerve, as well as a vessel network involving branches of the superficial temporal artery and posterior auricular artery with varying sizes of diameters are formed across the auricles[25,26]. This special and intricate subcutaneous structure may also provide rich physiological information varying with regions, such as BP, EEG and etc. However, most current ear-worn sensing devices with earplug-like or clip-like structures mainly focus on collecting data at a single location such as ear canal, earlobe, antihelix, etc. Thus, only temporal signal recording in specific region can be acquired and spatial-level characterization is missing. For instance, the pencil-like commercial auricular detecting tools, which are primarily single-probe electrical detectors (SPEDs) with a rigid metal probe (such as Pointer Excel II, Lhasa OMS Inc., Weymouth, MA, USA), can measure point-by-point cutaneous conductance levels by manually moving the probe over the auricular skin surface, but the acquired signal is extremely sensitive to applied pressure, leading to low measurement repeatability. Another emerging form of skin-integrated sensor is ultrathin flexible and stretchable electronics attached directly to the curved skin to monitor various physical signals[27-33]. However, it is also very challenging for these two-dimensional electrode arrays to cover the entire auricle and be reusable. Thus, there is a critical need for a conformable auricular sensing device that can provide a full-auricle measurement of physiological signals to investigate their spatiotemporal characteristics.

Here, we present the design and development of graphene-based distributed 3D sensing electrodes for the acquisition of spatiotemporal physiological signals of the human auricles. A novel shape-conformable personalized auricular sensor (3D-PAS) has been demonstrated to enable real-time electrophysiological signals mapping across the entire auricle. This platform, with printable and programable electrode threads, offers both conformable sensing interfaces with the curved auricular skin and 3D electrical interconnects. The entire sensor prototyping procedure, including human-specific auricular shape acquisition, 3D electrode pathways design and one-step multi-material 3D printing, is presented in this paper. Mechanical analysis of the skin surface curvature-dependent electrode sensing area design was also performed. As a proof-of-concept, simultaneous measurement of electrical skin resistance which may reflect the subcutaneous biological conditions and vascular or neural activities[34-39] at multiple auricular points (APs) was demonstrated. An auricular electrical skin resistance (AESR) contour is generated for 3D AESR mapping. For the first time, subject-specific AESR distributions and auricular region-specific AESR changes following physical exercise are demonstrated with unsupervised machine learning techniques. Finally, the correlations of AESR with HR and BP signals were also studied. These results provide potentially a universal platform and methodology for biomedical applications based on physiological signals of the auricles.

# RESULTS

**Design and fabrication.** The entire fabrication process included three steps. First, geometrical information on the human-specific outer ear structure was acquired and transferred (Fig. 1a). This began with the creation of a solid, 3D ear impression that shaped conformably with the auricle, whose surface curvature varies greatly between individuals and possesses a complex geometry. As shown in Fig. 1a(i), medical-grade molding polymers were evenly mixed and then filled into the entire outer ear, resulting in a personalized auricular mold (PAM). Structural light-based 3D scanning (Fig. 1a(ii)) was utilized on the PAM to generate a point cloud (Fig. 1a(iii)), which was later merged into a solid 3D-PAM geometry (Fig. 1a(iv)) in a 1:1 scaled form. Second, a geometric layout for the 3D electrode pathways was designed inside the acquired PAM model (Fig. 1b). Multiple electrodes were located at specific APs (Fig. 1b(i)), where spatially separated pathways were extended across the PAM model accordingly and embedded inside with a tortuous layout to serve as 3D interconnects (Fig. 1b(ii)). Here, the sensing areas of electrodes at multiple APs with varied surface curvatures were fixed to the same by scaling the geometric parameters (such as cross-sections, diameter, and orientation) of the 3D pathways (Fig. 1b(iii)). Third, the sensor was one-step prototyped (Fig. 1c). A well-designed PAM prototype incorporated with multiple electrode pathways, defining the overall format of the sensor, was subsequently rendered by a commercial dual-nozzle 3D printer integrating both a flexible elastomer and conductive graphene-enhanced polylactide (g-PLA) (Fig. 1c(i-iv)). The use of this materials combination provides good printability, comfortability and superior conductivity. The result of these steps was the creation of a 3D conformable and mechanically stable electrode-skin sensing interface. The hardware control units were then connected to the printed electrode channels for data acquisition. Herein, we delivered a universal strategy involving a wearable 3D-PAS device that offers a platform for geometrically integrating 3D-distributed and conformable electrodes to achieve simultaneous multi-region AESR collection in real time across the entire auricle.

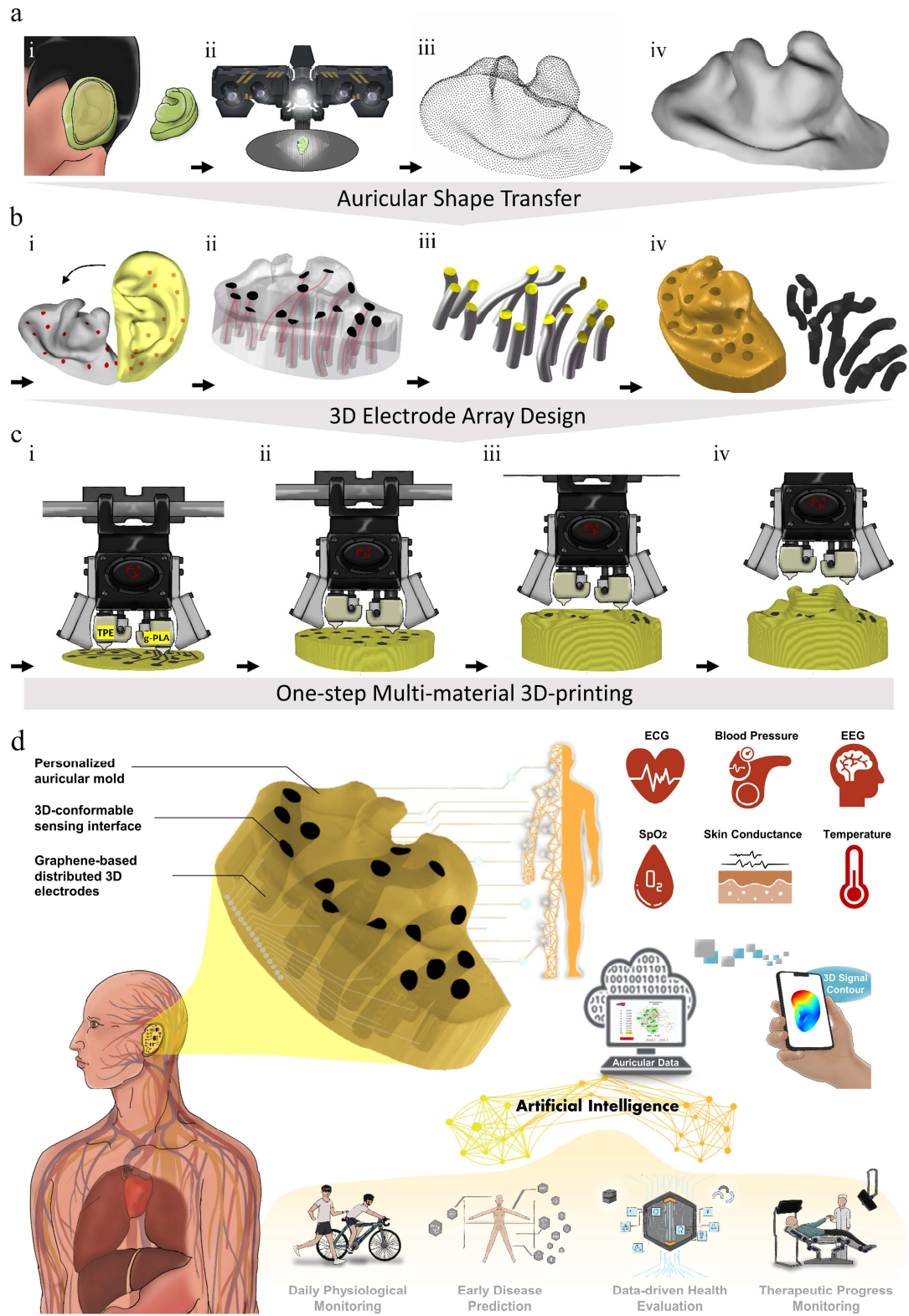

**Fig. 1. Diagram of the 3D-PAS design and fabrication process.** (a) Illustration of human-specific auricular skin-shape acquisition and transfer. i) Outer ear impression molding; ii) 3D scanning of the PAM; iii) point cloud generation; iv) CAD modeling of the PAM. (b) Illustration of 3D electrodes layout design. i) APs locating; ii) 3D electrode-pathways embedding; iii) geometric parameters scaling of 3D electrodes to achieve the same sensing area; iv) sensor-prototype assembling with 3D electrodes and the PAM. (c) Illustrations of one-step sensor prototyping. i-iv) sequential snapshots of 3D printing process integrating flexible TPE and conductive g-PLA materials. d) Schematic diagram of the 3D-PAS monitoring platform and data analysis route for potential biomedical healthcare applications.

**Mechanical analysis.** A critical feature of this device is that it can be conformably shaped to the entire curved auricle and geometrically configured with multiple 3D electrodes with a profile-controlled layout. The auricle of the ear, one of the smallest functional human organs, forms an ultra-curved skin surface (Fig. 2a). This curvature varies not only within an auricular region but also across individual subjects, and this is also reflected in the geometry of the PAM. From one of the cross-sections of the PAM (Fig. 2b), it can be observed that the curvatures of 10 points located along the cross-sectional curve differ up to tenfold (Fig. 2c). As a result, cylindrical electrode pathways with the same diameter across multiple APs with different surface curvatures and orientations will generate widely varying sensing areas (i.e., contact area) (Fig. 2d-e), which can cause great deviations in contact resistance. Therefore, once the cross-sections of the electrode pathways were determined, their geometric parameters (diameter, orientation, etc.) can be scaled point-by-point to achieve a consistent sensing area. Furthermore, wearing the 3D-PAS provides a 3D conformable and stable interface between the device and the entire auricle (Fig. 2f(i)). As a result, three repeated AESR measurements on each of human subjects demonstrated excellent repeatability of the sensor with only a 4.9% average coefficient of variation (CV) among all testing points (Fig. 2f(ii)). However, a commercial SPED had a much larger CV of approximately 35%, which was mainly caused by the significant deviation from operating pressure when moving across the auricle manually and freely without quantitative force feedback. Here, an overall surface curvature-dependent geometric design for the electrodes is implemented to ensure both conformability and consistency for the 3D AESR sensing interface.

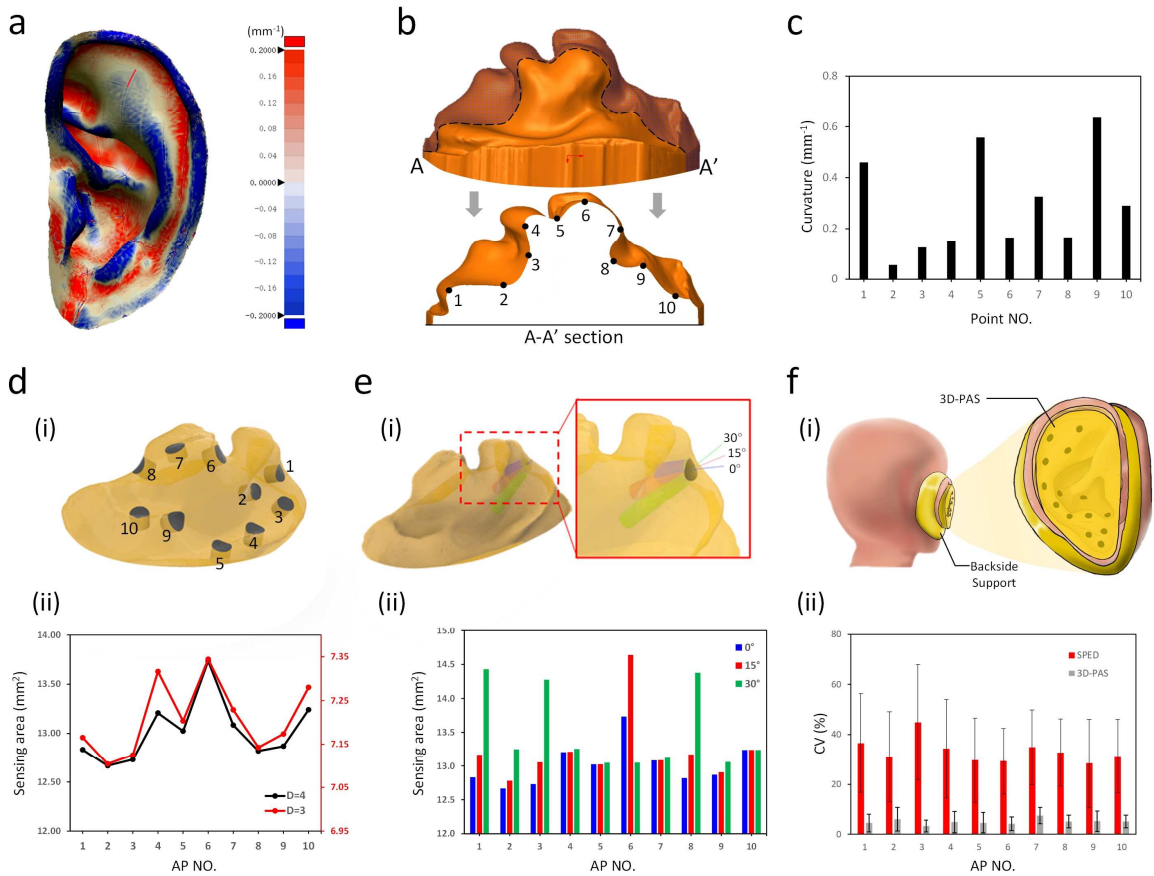

**Fig. 2. Mechanical analysis of the 3D-PAS.** (a) Overall curvature distribution across the entire human auricle. (b) Illustration of A-A' cross-section of 3D-PAM. (c) Curvature comparison at multiple points located along the cross-sectional curve. (d) Comparison of the electrode sensing area at multiple AP locations with varying surface curvatures. i) Illustration of cylindrical electrode pathways with the determined diameter across spatially distributed APs; ii) Sensing area comparison between multiple electrodes with pathway diameter D=3 and 4 mm. (e) Comparison of the electrode sensing area with different pathway orientations at each AP location. i) Illustration of electrode pathways with the same size (3 mm diameter) across the AP with different orientations (blue: normal to surface, red: 15° from normal, green: 30° from normal); ii) Sensing area comparison between different electrode pathway orientations at multiple AP locations. (f) Analysis of AESR measurement repeatability. i) Illustration of the mechanically stable sensing interface between the 3D-PAS and auricular skin; ii) Comparison of AESR measurement CV between a commercial SPED and the 3D-PAS.

**Spatial AESR mapping.** To demonstrate simultaneous multi-region AESR monitoring with the 3D-PAS, the positions of multiple testing points were first located (Fig. 3b). For each ear, a specific 3D-PAS device was prototyped by the abovementioned procedures for AESR signal collection with a data acquisition unit. The whole electrical loop includes a sensor, the auricular skin, body and LCR meter, where the circuit logically switches between the multiple electrodes with a multiplexer that prevents channel signal interference (Fig. 3a). The AESRs were directly read from a self-developed user interface in real time, which also marked all AP locations and highlighted the AP with the lowest AESR for further analysis; this can help systematically screen potential patients with abnormal AESR signals and advance auricular diagnoses. The AESR signals were stably collected across several minutes of monitoring (Fig. 3c). The variation in AESR levels across all testing APs in each measurement was drawn as a trend line (Fig. 3d), which maintains highly consistent among multiple tests indicating excellent repeatability of measurement. Furthermore, based on the AESR data spatially mapped at the scattered APs, a 3D AESR contour with both geometric and spatiotemporal signal information was generated for the first time by a numerical interpolation algorithm to visualize the overall AESR distribution across the entire auricle (Fig. 3e). This universal methodological route for 3D AESR mapping can be used to acquire quantifiable data for further study.

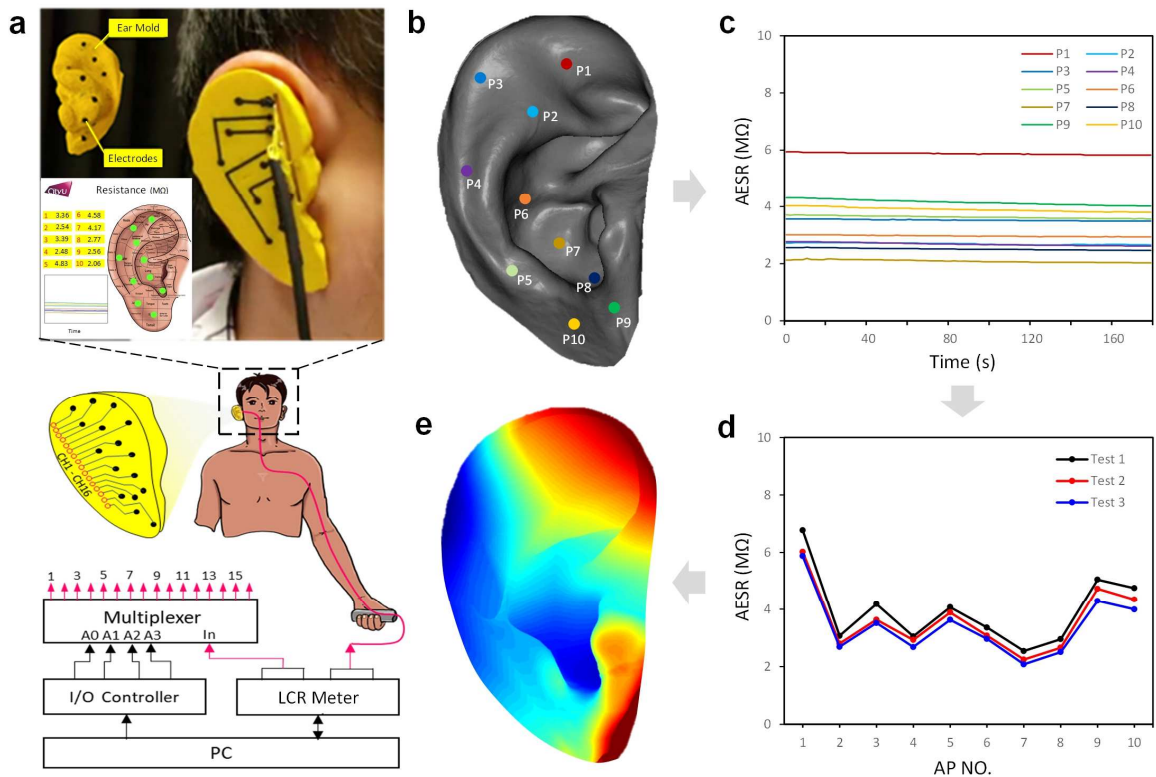

**Fig. 3. AESR mapping across the entire auricle.** (a) Illustration of AESR signal collection by the ear-worn 3D-PAS with a user interface and electrical control unit for multiplexed data acquisition. (b) Spatial distribution of AP locations across the entire auricle. (c) Simultaneously stable AESR signal collection at multiple APs in real time. (d) Highly consistent AESR variation across all tested APs between repeatable tests. (e) 3D AESR contour generation for visualizing the overall AESR distribution across the entire auricle.

**Human-specific AESR distribution.** Using 3D-PAS tools, comprehensive studies on the characteristics of the distribution of AESR among a human population were performed. Here, personalized devices were prototyped for both the left and right ears of 30 human subjects. In this paper, 10 APs across the entire auricle were selected and located on all ear subjects with a proportionally scaled layout (Fig. 4a). After placing the 3D-PAS into the ear, AESR signals were collected with the same procedures from all subjects, followed by data normalization. The AESR distribution across each ear was then characterized as a trend line which is drawn by connecting all the normalized AESRs of 10Aps and packaged as a single dataset (i.e., a matrix of 1 by 10). Unsupervised machine learning techniques, including principal component analysis (PCA) and K-means clustering, were performed on the AESR datasets collected from all 60 ears (i.e., a pair ear per subject). Firstly, PCA achieved 86.1% of the total explained variance (EV) from raw data with 3 principal components (PCs). Then, K-means clustering algorithm was performed, here we calculated the sum of the square error (SSE) and use the elbow method to find that 4 was the optimal value of cluster number K that yielded the greatest cluster-to-cluster separation and best within-cluster gathering; we also achieved a high silhouette score (S-score) of 0.76. This result indicates that all AESR datasets were clearly classified into 4 clusters, while each cluster showed a specific AESR distribution across the 10 APs. Accordingly, 4 clusters of 3D scattered points were acquired by PCA dimensionality reduction, and were plotted to visualize the clustering result (Fig. 4b). Four trend lines with shaded error bands (Fig. 4d) also depicted the specific AESR distribution of clusters A, B, C and D, which included 35, 17, 5 and 3 ears, respectively (Fig. 4c). By matching the clustered datasets with the subject labels, it was observed that 80% of the tested subjects (i.e., 24/30) had their left- and right-ear AESR distribution classified into the same cluster (Fig. 4e). This preliminarily indicates that AESR distribution information of both auricles is highly similar. Furthermore, the 3D AESR contours of all ears were generated; the clustering result can also be characterized by these contours, which directly illustrate the high consistency in the 3D AESR distribution among within-cluster subjects as well as the large variability among across-cluster subjects (Fig. 4f). Here, we demonstrated a novel biometrical tool for human population classification, which shows great potential to serve as a new form of biological marker.

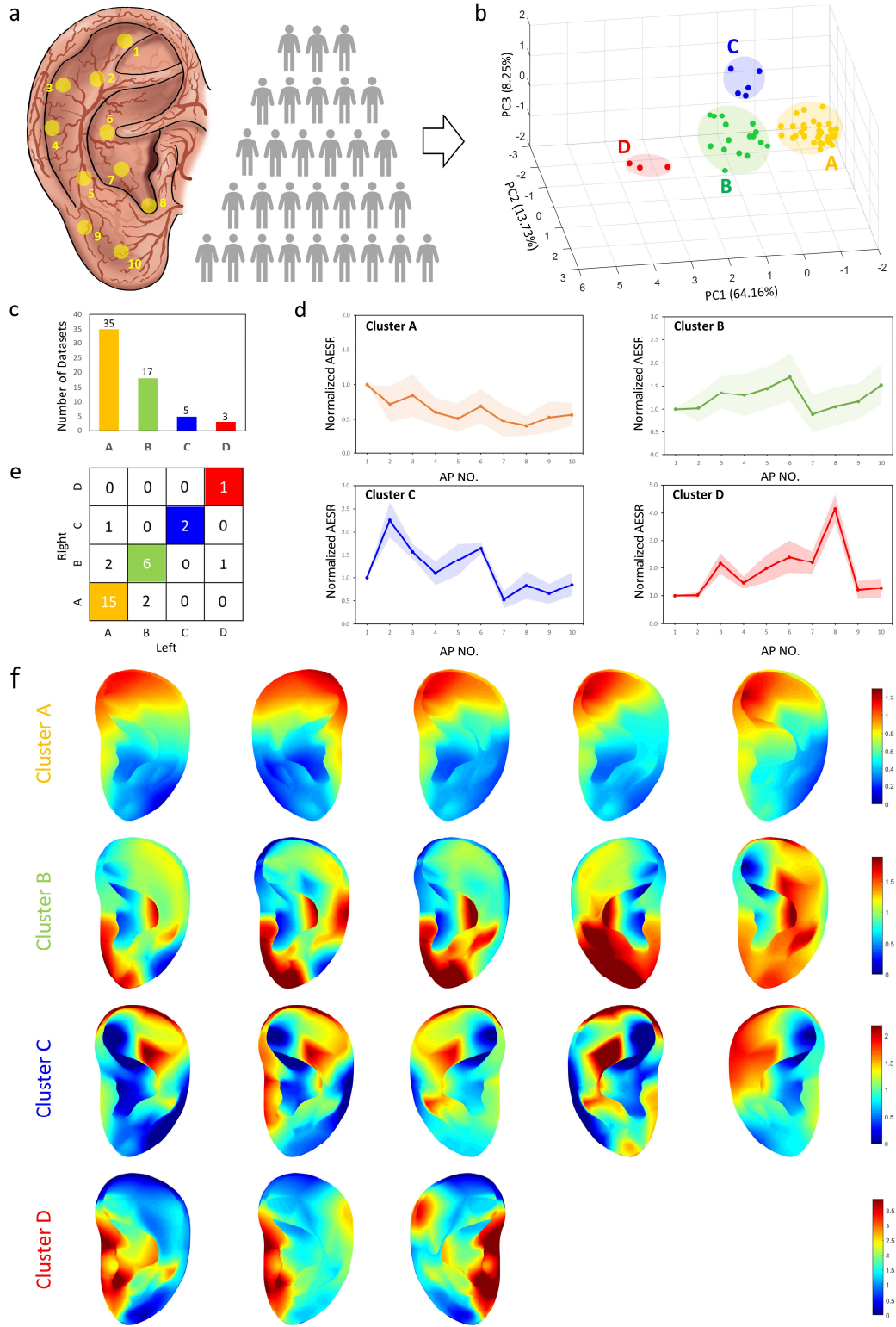

**Fig. 4. Human-specific AESR distribution analysis.** (a) Schematic illustration of 10 AP locations for all human auricles. (b) K-means clustering result for the AESR datasets collected from all 60 ears shown on a 3D scatter plot (each point denotes one AESR dataset from a single ear after PCA). (c) Distribution of the total ear subject number in each cluster. (d) Matching results of both the left- and right- ear datasets from each human subject after K-means clustering (horizontal: cluster label of the left ear, vertical: cluster label of the right ear; diagonal numbers denote the number of subjects for whom both the left and right AESR datasets were classified to the same cluster (A, B, C or D). (e) Specific AESR variation trendlines across all tested APs with shaded error bands for each cluster. (f) 3D AESR contours visualization for specific AESR distribution of ear subjects from each cluster (5 representative subjects are selected in clusters A and B).

**Region-specific AESR change after exercise.** Using 3D-PAS tools, the study on individual AESR change when subjected to physiological stimuli was investigated for the first time. Stationary cycling, a popular physical exercise worldwide, causes a whole-body physiological response[40], including an increase heart rate (HR) to meet the muscles' demand for oxygen and a surge in blood pressure (BP) with the increased availability of vasodilatory mediators. To explore the effects of physical exercise on AESR levels, 17 volunteers participated in a study to repeat stationary cycling exercises three times at a fixed intensity (i.e., test A1, A2 and A3). Here, 3 additional APs (i.e., AP3, 7 and 11) were added to the 10 previously studied APs (Fig. 5f). All tests followed the same procedures. During each test, biosignals were collected at 4 fixed periods (i.e., period I, II, III and IV) before and after cycling (Fig. 5a). Two additional tests of controls were also performed without cycling (i.e., test B1 and B2), where the AESR signals were collected in the same 4 periods. Meanwhile, HR and BP were recorded by a commercial smart watch and a cuff-style BP monitor, respectively. Five minutes after the completion of the exercise, it was observed that HR and BP increased by 42.9% and 16.0%, respectively, on average over all tests (Fig. 5b-c). Correspondingly, the AESR levels at AP1-6 dramatically dropped by 60.8%, 66.8%, 55.4%, 64.9%, 57.8% and 51.3% on average, while those at AP7-13 dropped by less than 23.5% (Fig. 5d). In 98% of total 51 individual cycling tests, auricular region-specific AESR changes were observed under the stimuli of physical exercise. Comparatively, AESR levels remained no change across the four periods in all tests of controls. Similar change of the auricular electrical skin impedance (AESI) scanning from 4 to 4000 Hz at AP1-6 was also shown in Fig. 5e.

For each measurement, the AESRs collected at each of the 13 APs were point-to-point normalized by their initial AESRs (AESR I) and repackaged into a new dataset (i.e., matrix of 1 by 13) labeled as 'Subject-Test-Period'. Unsupervised machine learning was then utilized to classify the datasets comprising the data from all five tests for each volunteer, including three datasets from the repeated cycling tests and two from the control tests, using the AESR II. Following PCA which achieved a 98.5% average total EV with 3 PCs, K-means clustering was performed, the optimal value of cluster number

K determined by the elbow method was 2, while the S-score was 0.88 on average, which indicates that the two clusters of datasets can be clearly classified. In the analysis for 94% (i.e., 16 out of 17) of the volunteers, datasets in these two clusters were exactly label-matched to the AESR signals collected in cycling tests and control tests, respectively. A 3D scatter plot was shown to visualize the individual clustering results (Fig. 5g(i)). Furthermore, similar PCA and clustering analyses were performed for the datasets comprising the data from all the tests on 17 volunteers. Following PCA, the total EV with 3 PCs was 88.3%, and similar to the prior analysis, two clusters (Fig. 5g(ii)) were classified with K=2 and an S-score of 0.85.

Additionally, it was also observed that the AESR at periods III and IV of AP1-6 gradually increased back to 67.7% and 100% of the initial value on average, demonstrating a highly consistent trend with HR and BP. Here, unsupervised machine learning was also used to classify the normalized datasets comprising the AESR data collected at the 4 periods across the three repeated cycling tests for each volunteer. First, PCA achieved 93.4% total EV on average with 3 PCs. Then K-means clustering was performed; when K was determined as 3, analysis for datasets from 76.5% (i.e., 13 out of 17) of the volunteers obtained the result that, one dataset of "AESR II" together with three datasets of "AESR IV" were classified into cluster A, three datasets of "AESR II" were classified into cluster B, and three datasets of "AESR III" were classified into cluster C. A 3D scatter plot visualized this clustering result, in which three arrows were drawn to illustrate cluster-point shifting and match the dynamic evolution of the whole-body response (Fig. 5h(i)). Furthermore, similar machine learning results were obtained for the AESR datasets collected from all cycling tests from the 17 volunteers; PCA yielded 82% total EV with 3 PCs, the elbow method yielded a K of 3, the S-score was 0.64, and 85.1% of the datasets achieved the same label-group matching result described in the above individual volunteer analysis (Fig. 5h(ii)). Last, a highly consistent region-specific color change across the four periods can be observed from the generated 3D AESR contours (Fig. 5i). Here, both data-driven and vision-driven analyses validate the auricular region-specific AESR change as the response to physical exercise stimuli.

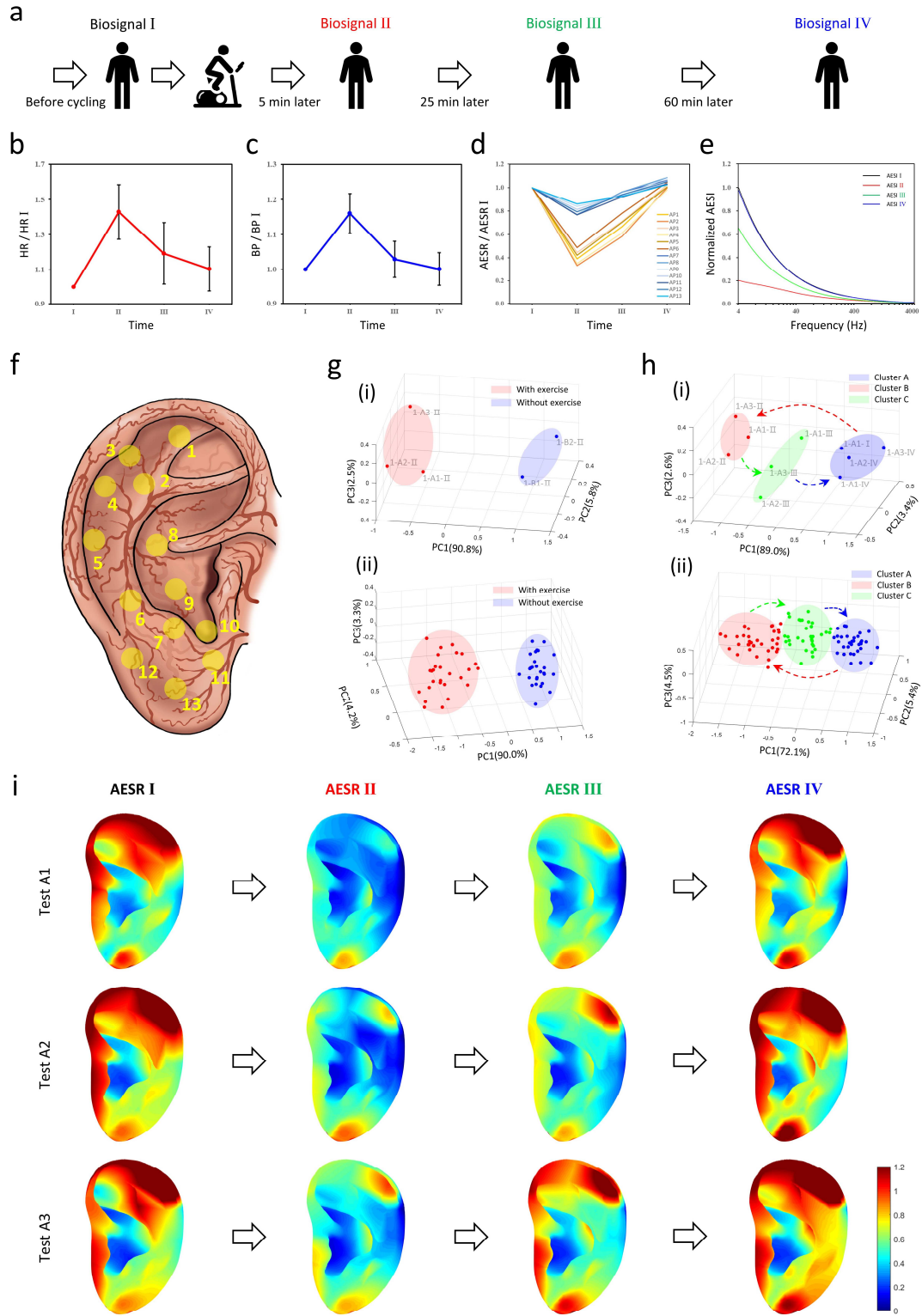

**Fig. 5. Analysis of after-exercise region-specific AESR change.** (a) Schematic illustration of biosignals (HR, BP, AESR and AESI) collection in four sequential periods in each cycling test. (b) HR, (c) BP, (d) AESR, and (e) AESI changes after fixed-intensity cycling exercise. (f) Locations of the studied 13 APs for all human ear auricles. (g) 3D scatter plots showing K-means clustering results for the AESR datasets collected from the cycling/control tests for i) an individual volunteer and ii) all 17 volunteers. ("1-A1-II" denotes the AESR dataset collected in the period II in the first cycling test from volunteer 1) (h) 3D scatter plots showing K-means clustering results for the AESR datasets collected in each of the four periods from the three cycling tests for i) an individual volunteer and ii) all 17 volunteers. (i) 3D AESR contours illustrating overall AESR distribution changes across all four periods in the cycling tests (one representative volunteer is selected).

Furthermore, it was observed that AESR changes differed greatly test-to-test after fixed-intensity exercise. This indicated the presence of individual body condition differences, which were also reflected in the changes in HR and BP signals. Scatter plots comparing the changes in AESR with the changes in HR and BP across all cycling tests from 17 volunteers are shown in Fig. 6. Statistical analysis was performed on these biosignals point by point. It obtained that the Pearson correlation coefficient (PCC) was 0.57 on average for the "AESR-HR" comparison across AP1-6, and was 0.47 on average for the "AESR-BP" comparison across AP1-5, indicating that AESR certainly correlates with HR and BP at most of the "active points" (AP1-6). On the other hand, there is no correlation of AESR with HR and BP ($P>0.05$ or $PCC<0.4$) at most of "inactive points" (AP7-13).

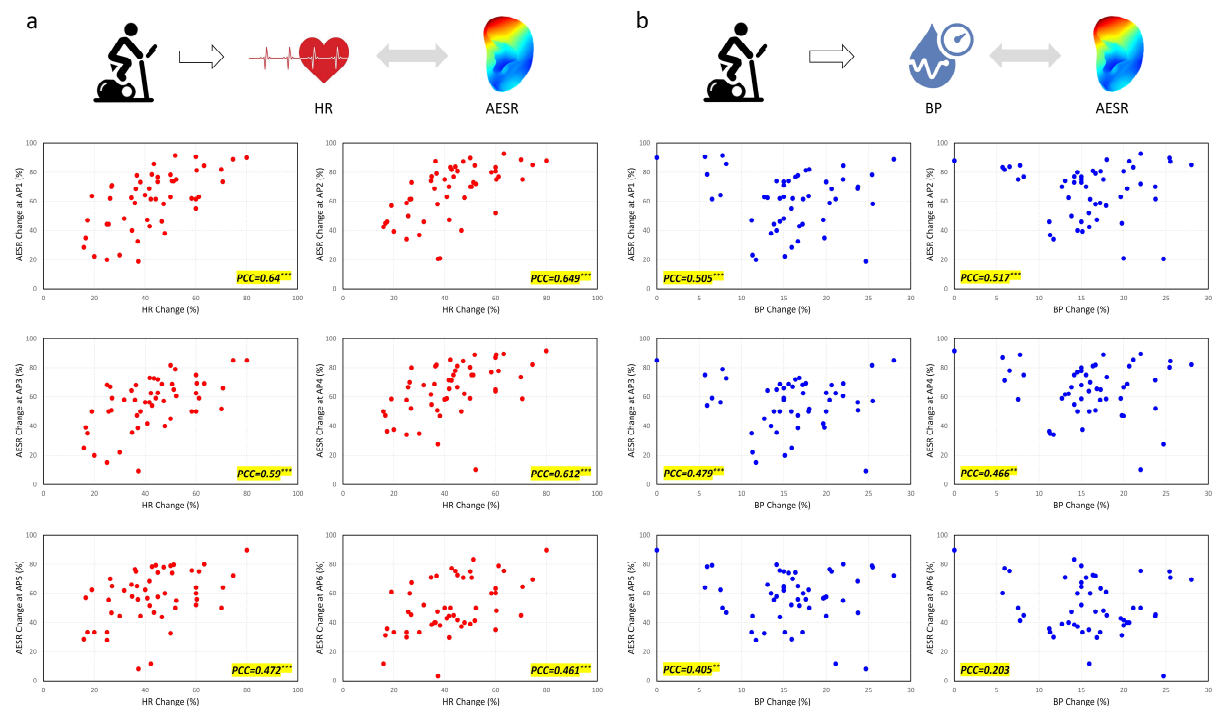

**Fig. 6. Correlation analysis of AESR with HR and BP.** Scatter plots of changes in AESR at AP1-6 versus changes in (a) HR and (b) BP after fixed-intensity cycling exercise (Biosignals II collected in all cycling tests from 17 volunteers are used).

## DISCUSSION

The design of the 3D-PAS provides a 3D conformable sensing interface with multiple electrodes whose locations, layout, size and density can be scalably altered by demand. In addition to AESR, other auricular physiological biosignals (such as BP, blood oxygen saturation, temperature, hydration) can also be monitored and mapped through the strategy of sensor design and data analysis introduced in this paper. Furthermore, diverse physical stimulating functions (electrical, mechanical, thermal, etc.) can be developed to assist auricular therapies by applying alternative functional materials for 3D electrodes. Accordingly, multimodal sensing together with stimulation over the entire auricle would be quite promising for comprehensive health evaluation and management.

In this paper, the characteristics of the AESR distribution were investigated among 30 human subjects. The result is that 4 types of AESR distributions across all testing APs were clearly classified. Interestingly, it was observed that subjects in cluster A, B and D have a recognizable age-level difference, which may reveal that the separation of these clusters can partially be explained by "age differences". However, given the age-level overlap for a few subjects from different clusters, additional hidden classification criteria may also exist. Therefore, more studies involving a larger cohort of subjects in the future may help in forming a comprehensive understanding of the underlying physical meaning for these interesting observations. Furthermore, within each cluster, the AESR distributions of all subjects were essentially similar with only minor differences between them. Meanwhile, the AESR datasets collected from repeated measurements on multiple subjects from the same cluster can also be clustered into individuals. This indicates that each subject has a unique form of AESR distribution, which, in the vein of a "fingerprint", can be called an "earprint". Given the above analyses, AESR signals hold great potential to serve as a new form of biological marker, similar to BP or HR, for matching individuals with diverse body conditions.

In the cycling tests on 17 volunteers, much greater AESR changes were consistently observed after exercise at AP1-6 than at AP7-13, which may reflect the varying levels of hemodynamic and neural activity underneath skin in different auricular regions. The gradual increase in the AESRs back to their baseline levels is consistent with the evolution of the condition of the body during recovery after exercise. In the statistical analysis, the fact that AESR didn't correlate with BP at AP6 and that smaller PCCs were obtained for the "AESR-BP" comparison than for the "AESR-HR" comparison may have resulted primarily from the high inconsistency in the noncontinuous BP measurement with the commercial, cuff-style BP monitor. Furthermore, approximately 10% of all the datasets were abnormal, 80% of which were collected from only two individual volunteers who may differ from the others in body condition. These findings provide quantitative evidence for auricular region-specific AESR change under physiological stimuli. Thus, following further studies, AESR monitoring is promising for personalized healthcare applications.

In summary, a personalized wearable device incorporating distributed 3D electrodes for simultaneous, multi-region AESR monitoring across the entire auricle was developed. Data analyses, including 3D contours for spatiotemporal AESR mapping and unsupervised machine learning for AESR datasets classification, reveal for the first time human-specific AESR distribution, and the correlations of AESR spatiotemporal electrophysiological signals with HR and BP. This strategy of 3D electrodes design and full-auricle sensing platform shows promising biomedical applications for spatiotemporal auricular physiological monitoring during daily activities.

## METHODS

**Fabrication of the 3D-PAS:** The whole process starts with the construction of personalized 3D auricular impressions by molding with medical-grade polymers and catalyzers (Green Eco, DETAX), which are mixed at a ratio of 1:1 and then injected into the entire auricle as well as the backside with a syringe. After a few minutes, the mixture solidifies, and the PAM is shaped. 3D scanning of the PAM is performed by a structural light-based 3D scanner (Dual-laser-source, 0.001 mm resolution, Nanyangmengyang Machinery Co., Ltd, China) to generate a point cloud, which is then merged into a 3D solid model whose surface is further modified in Computer-Aided Design (CAD) software. Testing points are located with a proportionally scaled layout. The layout, orientation and size of 3D electrode pathways are geometrically designed within the PAM in CAD software. A single sensor that incorporates the PAM and the 3D electrodes is then one-step prototyped with a dual-nozzle 3D printer (RAISE 3D PRO 2, USA), which integrates thermoplastic elastomer (TPE) (TPE-85A, Esun Industrial Co., Ltd., China) and g-PLA (BLACKMAGIC3D Company, USA) printable materials. The flexible TPE enables good comfortability in contact with the auricular skin and integrates well with the g-PLA material, offering a superior conductivity of 1.7 S·cm−1. The good printability of this materials combination is realized by optimizing the printing parameters (e.g., extruder temperature and extrusion rate). Last, commercial conductive gel is thinly coated onto the printed electrode surfaces prior to measurement.

**Measurement Calibration:** For each single electrode on the 3D-PAS, the accuracy of the electrical resistance measurement was calibrated by a series of constant resistors ranging from 0.5 to 10 MΩ. The temperature coefficient of resistance was also characterized in a laboratory oven as around 2.6 Ω/°C over a range of room temperatures (25-35°C), which is negligible relative to skin resistance. Therefore, the 3D-PAS can achieve good measurement accuracy, efficiency, visuality and repeatability far beyond those of manually operated, commercial SPEDs.

**Mechanical analysis:** The distribution of the curvature of the entire auricle skin surface is acquired by the internal curvature analysis function built in CAD software, which also provides the curvature-dependent sensing area for the conformable electrodes with predetermined locations and geometrical parameters.

**Human studies:** In total, 30 volunteers participated in the AESR distribution study, and 17 volunteers participated in the cycling exercise tests. Before each measurement, medical alcohol pads were used to remove oil, sweat and other impurities on the auricular skin surface followed by air drying. In the investigation of AESR change after physical exercise, each volunteer was asked to repeat three stationary cycling tests at a fixed intensity (7 km within 20 min) with at least a 48-hours interval on a fitness bike (Decathlon EB 500 SP, France) and two control tests without cycling during the same time slot on a separate day. For the individual tests, AESR, HR and BP signals were recorded in four sequential periods: before exercise (period I) and 5 min (period II), 25 min (period III) and 60 min after the completion of the exercise (period IV). Room temperature and humidity were also monitored, demonstrating negligible fluctuations during the whole test.

**Data acquisition:** *AESR* was measured by the LCR-meter function embedded in the impedance analyzer (HIOKI-IM3570, Japan), with a 16-channel multiplexer (ADG706) switching between multiple electrodes. **HR** was monitored by a commercial smart watch (HUAWEI WATCH GT 2 Pro, China). **BP** was measured by a cuff-style commercial monitor (Omron HEM-7320, Japan). ***Room temperature and humidity*** were measured by a commercial monitor (Jiandarenke COS-03, China).

**Data analysis:** ***AESR normalization.*** In the investigation of AESR distribution among a human population and generation of 3D AESR contours, AESR levels measured at all testing APs from each subject were normalized to those at AP1 for further analysis. In the investigation of AESR change under physical exercise conditions, including both cycling and control tests, the AESRs collected in periods I to IV at each AP were normalized to those in period I (i.e., initial value). ***3D AESR contour generation.*** The normalized AESR signals collected at multiple APs and 3D coordinates of the point cloud acquired from auricular shape 3D-scanning were input to perform natural neighbor interpolation for 3D scattered data in MATLAB, generating a 3D AESR contour with a continuous AESR gradient to visualize the overall spatial AESR distribution across the entire auricle. ***AESR dataset package.*** Normalized AESRs collected at *N* APs in one single measurement were packaged into a single dataset, that is, a matrix of 1 by *N*. ***Unsupervised machine learning analysis.*** All normalized AESR datasets collected from *M* tests were incorporated into an input matrix (*M\*N*), which was later dimension-reduced to a *M\*3* matrix by PCA. Then, the K-means clustering algorithm was utilized to classify the datasets in Matlab, where the Euclidean distance and "K-means++" clustering center initialization were used; the other parameters were set to their default values. ***Determination of K.*** The optimal value of cluster number K was determined by the "elbow method", that is, by selecting the elbow point in the SSE-K plot. The SSE for each cluster is calculated as Eq. (1) and the S-score of each dataset was acquired as Eq. (2). ***Correlation analysis.*** P-values for significance testing and PCC were acquired by SPSS software after excluding abnormal datasets.

$$SSE = \sum_{i=1}^{k} \sum_{x \in C_i}^{t} dist^2(m_i, x) \quad (1)$$

where $x$ is a data point in cluster $C_i$ and $m_i$ is the center point for cluster $C_i$.

$$s(i) = \begin{cases} 1 - a(i)/b(i), & \text{if } a(i) < b(i) \\ 0, & \text{if } a(i) = b(i) \\ b(i)/a(i) - 1, & \text{if } a(i) > b(i) \end{cases} \quad (2)$$

where $a(i) = \frac{1}{|C_i|-1}\sum_{j \in C_i, i \neq j} d(i,j)$, $b(i) = \min_{k \neq i} \frac{1}{|C_k|}\sum_{j \in C_k} d(i,j)$, and i is a data point in cluster $C_i$.

## ACKNOWLEDGEMENTS

We acknowledge the support of the Hong Kong Research Grants Council's Joint Laboratory Funding Scheme (Project No.: JLFS/E-104/18), the Hong Kong Research Grants Council's General Research Fund (Project No.: 11204918) and the University of Hong Kong Seed Fund for Translational and Applied Research (Grant No.: 201711160034).

## ETHICAL DECLARATION

All human cycling studies were implemented in accordance with the ethical guidelines and with the approval of the Human Subject Ethics Committee of City University of Hong Kong (Reference No.: 10-2020-19-F). Written informed consent was obtained from all participants with the purpose and procedures of the study clearly explained.

## AUTHOR CONTRIBUTIONS

W.J.L., Q.H. and H.Y.C. conceived of the basic idea of this study and sensor design. W.J.L., Q.H., and C.W. designed the experiments. Q.H. performed the experiments. Q.H., W.J.L., H.Y.C., X.Y., V.A.L., and Y.S. analyzed the data. M.Y. and L.L. provided support for medical related information and interpretation of data for this study. K.Y., J.L., H.S. and S.H. assisted in part of experiments and sensor fabrication process. Q.H., W.J.L. and C.W. co-drafted this paper. X.Y., V.A.L. and Y.S. provided critical technical advice for this study. W.J.L. and H.Y.C. supervised the project.